\documentclass[a4paper,11pt]{jpconf}
\usepackage{graphicx}
\usepackage{epstopdf}
\usepackage{graphicx}
\usepackage{dcolumn}
\usepackage{bm}
\usepackage{amsmath}
\usepackage{color}
\usepackage{ulem}

\newcommand{\be}{\begin{equation}}
\newcommand{\ee}{\end{equation}}
\newcommand{\beq}{\begin{eqnarray}}
\newcommand{\eeq}{\end{eqnarray}}

\def\nue{\mathrel{{\nu_e}}}
\def\numu{\mathrel{{\nu_\mu}}}
\def\nutau{\mathrel{{\nu_\tau}}}

\def\barnumu{\mathrel{{\bar \nu}_\mu}}

\def \gta {\mathrel{\vcenter{\hbox{$>$}\nointerlineskip\hbox{$\sim$}}}}

\def\t13{\mathrel{{\theta_{13}}}}
\def\y12{\mathrel{{\tan^2 \theta_{12}}}}
\def\c2{\mathrel{{\chi^2 }}}

\newcommand{\n}{neutrino}

\newcommand{\fb}{FB}
\newcommand{\ic}{IceCube}

\begin{document}

\title{\small{Proceedings of The 59th Annual Conference of the SA Institute of Physics, 2014}
\\
\Large{IceCube Neutrino Events from Fermi Bubbles}}

\author{C Lunardini$^1$, S Razzaque$^2$, and L Yang$^{1,2,3}$}

\address{$^1$ Department of Physics, Arizona State University, Tempe, AZ 85287-1504, US}
\address{$^2$ Department of Physics, University of Johannesburg, PO Box
  524, Auckland Park 2006, South Africa}
\address{$^3$ Laboratory for Astroparticle Physics, University of Nova Gorica, Vipavska 13, 5000 Nova Gorica, Slovenia}

\ead{lili.yang@ung.si}

\begin{abstract}
The IceCube Neutrino Observatory announced thirty-seven candidate events observed with deposited energies above $\sim$ 30 TeV with three-year dataset, more than expected from atmospheric backgrounds. We discuss the detectability of the Fermi Bubbles (FB) by IceCube and show that up to 6 -- 7 of the 37 events could originate from the FB. If the observed gamma rays from the FB are created due to baryonic mechanism, high-energy ($>$ GeV) neutrinos should be emitted as a counterpart. These neutrinos should be detectable as shower- or track-like events at a Km$^3$ neutrino detector. For a hard primary cosmic-ray proton spectrum, $E^{-2.1}$, and cutoff energy at or above 10 PeV, the FB flux substantially exceeds the atmospheric backgrounds. For a steeper spectrum, $E^{-2.3}$, and/or lower cutoff energy, detection with high significance will require a longer running time.
\end{abstract}

\section{Introduction}
The IceCube Neutrino Observatory recently found strong evidence for high energy astrophysical neutrino flux at the level of $10^{-8}$ GeV/cm$^2$/s/sr in three yeas of data \cite{Aartsen:2014gkd}. Except the first 28 neutrinos announced before \cite{Aartsen:2013jdh}, an additional 9 neutrino candidates from the third year of data has been reported \cite{Aartsen:2014gkd}. These total 37 events with observed energy of 30 TeV -- 2 PeV more than expected background of 15 events from atmospheric muons and neutrinos, indicate a 5.7 $\sigma$ rejection of purely atmospheric hypothesis. Of these events, 9 are identified with visible muon tracks. The remaining 28 events are cascades or showers, caused by neutrino interactions other than $\nu_{\mu}$ charged current. 

The origin of these high energy neutrinos at IceCube is still mysterious \cite{Aartsen:2014cva}. Although the prompt atmospheric neutrino may contribute significantly with energy between 10 and 100 TeV \cite{Lipari:2013taa,Aartsen:2014muf}, the isotropic distribution of these events indicates astrophysical sources are the most natural explanation. Many scenarios have been discussed in the literature, including cores of active galactic nuclei \cite{Stecker:2013fxa,Winter:2013cla}, $\gamma$-ray bursts \cite{Razzaque:2013dsa} and their lower-powered counterparts \cite{Razzaque:2003uv, Murase:2013ffa}, galaxies with intense star-formation \cite{Loeb:2006tw, Stecker:2006vz, Liu:2013wia}, active galaxies \cite{Kalashev:2013vba}, flat-spectrum radio quasars \cite{Dermer:2014vaa} and intergalactic shocks \cite{Murase:2013rfa}. Due to the weak cluster of events near the Galactic Centre, some analyses implied that Galactic sources may be responsible for a fraction of the signals \cite{Fox:2013oza, Gonzalez-Garcia:2013iha}. The \ic\ events may originate from the Galactic Centre (GC) \cite{Razzaque:2013uoa}, the \fb\ \cite{Razzaque:2013uoa, Ahlers:2013xia}, Galactic Halo \cite{Taylor:2014hya}, or Galactic plane \cite{Anchordoqui:2013qsi,Joshi:2013aua}. More studies suggested that the origination could also be from PeV dark matter decay \cite{Feldstein:2013kka,Esmaili:2013gha, Barger:2013pla}.

In this paper, we investigate the possibility that \ic\ might be observing signals from the \fb. The two gamma-ray bubbles, observed by Fermi-LAT \cite{Su:2010qj}, are symmetrically located above and below the Galactic centre, extending up to $50^{\circ}$ in Galactic latitude. The origin of the \fb\ is still unknown. There are two main hypothesis, the supermassive black hole activity \cite{Su:2010qj} or high rate of star formation in the GC \cite{Crocker:2010dg, Fujita:2013jda}. The emitted gamma rays from the \fb\ are produced either via leptonic model, which is the Inverse Compton scattering of low energy photons by highly relativistic electrons, or via hadronic model, i.e., decays of $\pi^{0}$ created from interactions of energetic baryons with gas in the \fb. In the hadronic model, a neutrino counterpart with similar magnitude as gamma ray flux from the \fb\ is produced \cite{Crocker:2010dg, Lunardini:2011br}, which could be detected as muon tracks or showers at KM3NeT, ANTARES, or \ic\ \cite{Adrian-Martinez:2013xda,Adrian-Martinez:2012qpa,Lunardini:2011br}. The ANTARES collaboration has placed an upper limit for this neutrino flux, as seen in Fig.~\ref{fluxfig} \cite{Adrian-Martinez:2013xda}. Later, we will quantitatively present the down-going muon track as well as shower events from the \fb\ at \ic\ , not only as a possible interpretation of \ic\ events, but also as prediction for future searches. 

\section{Neutrino Fluxes}

\begin{figure}[h]
\begin{center}
\begin{minipage}{17pc}
\includegraphics[width=17pc]{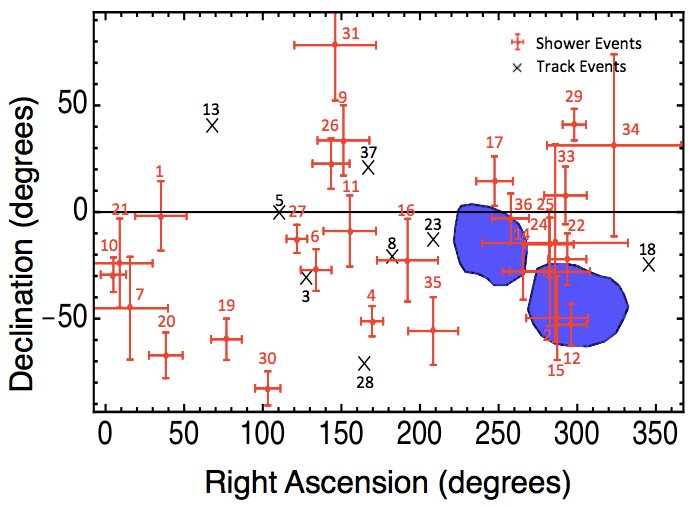}
\caption{\label{icdata}\ic\ events distribution in equatorial coordinates, with their median angular errors, from \cite{Aartsen:2013jdh}. The corresponding FB regions are shown with shaded contours. }
\end{minipage}\hspace{3pc}%
\begin{minipage}{17pc}
\includegraphics[width=17pc]{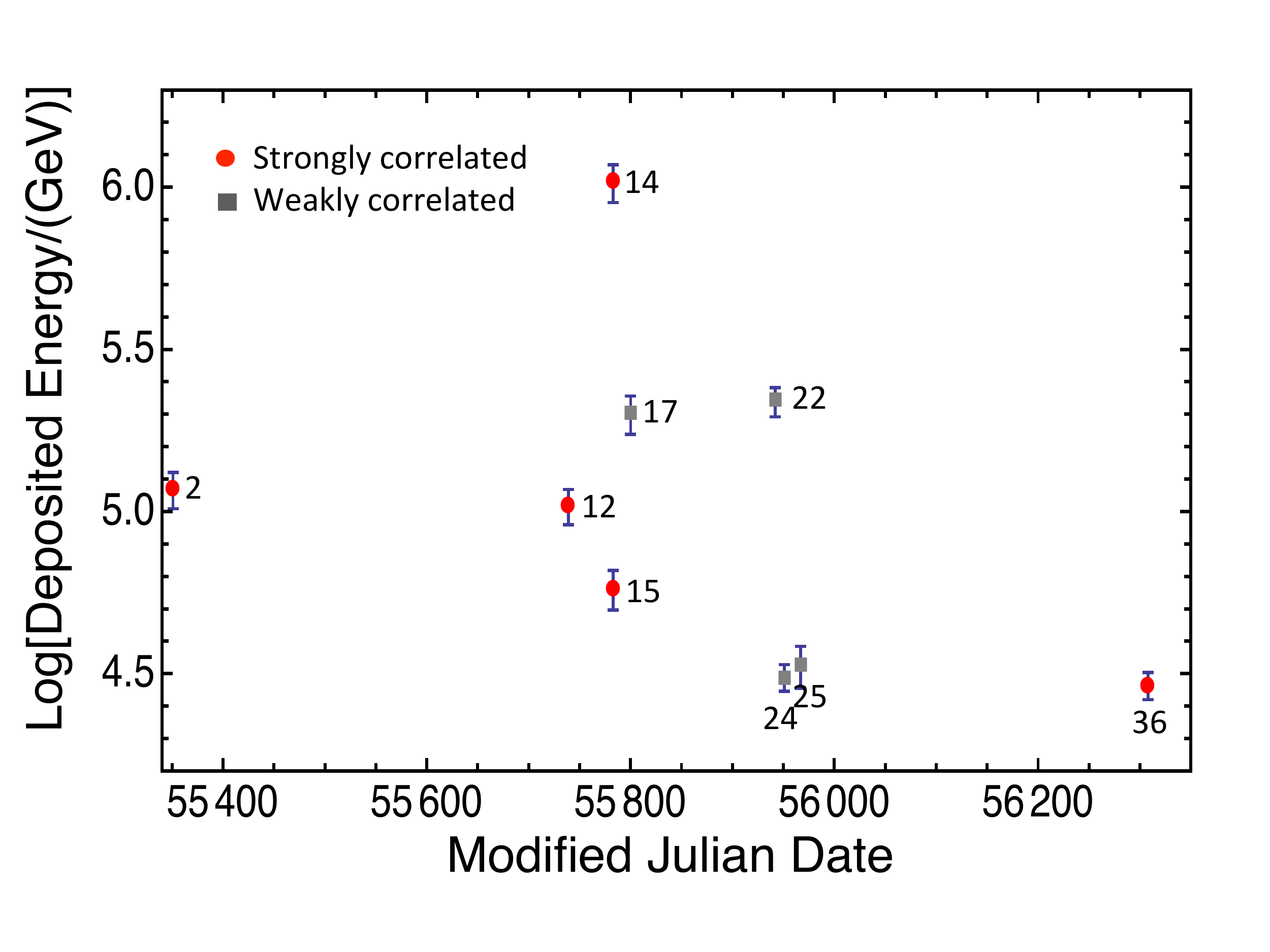}
\caption{\label{ictime}Events 2, 12, 14, 15, 17, 22, 24, 25, 36 (spatially correlated with bubble) are shown with errors on energy in modified Julian time and deposited energy distribution.}
\end{minipage} 
\end{center}
\end{figure}

The Fermi Bubbles are seen as extended sources in the southern sky, see Fig.~\ref{icdata}, subtending a 0.808 sr solid angle \cite{Su:2010qj}. Their gamma ray flux is fairly uniform over the extent of the bubbles \cite{Su:2010qj}, and it is expected the neutrino flux has the same feature \cite{Lunardini:2011br}. We compare the coordinates of the 37 \ic\ all-sky search events and their median angular errors \cite{Aartsen:2013jdh} with the bubbles coordinates, see Fig.~\ref{icdata}. Note that event 32 produced by pair of background muons is not labeled in Fig.~\ref{icdata}, because it cannot be reconstructed with a single direction and energy \cite{Aartsen:2014gkd}. It appears at least 5 events, number 2, 12, 14, 15, 36 are strongly correlated with the \fb\, whose central position values are inside the bubbles. In addition, events 17, 22, 24,25, are weakly correlated with the \fb\ whose values, within errors, are compatible with the bubbles. Among these 9 correlated events, event 14, has $\sim$ 1 PeV deposited energy \cite{Aartsen:2013jdh}. Fig.~\ref{ictime} displays the time correlation of these 8 events with errors on the observed energy.

\begin{figure}[htbp]
\centering
\includegraphics[width=0.5\textwidth]{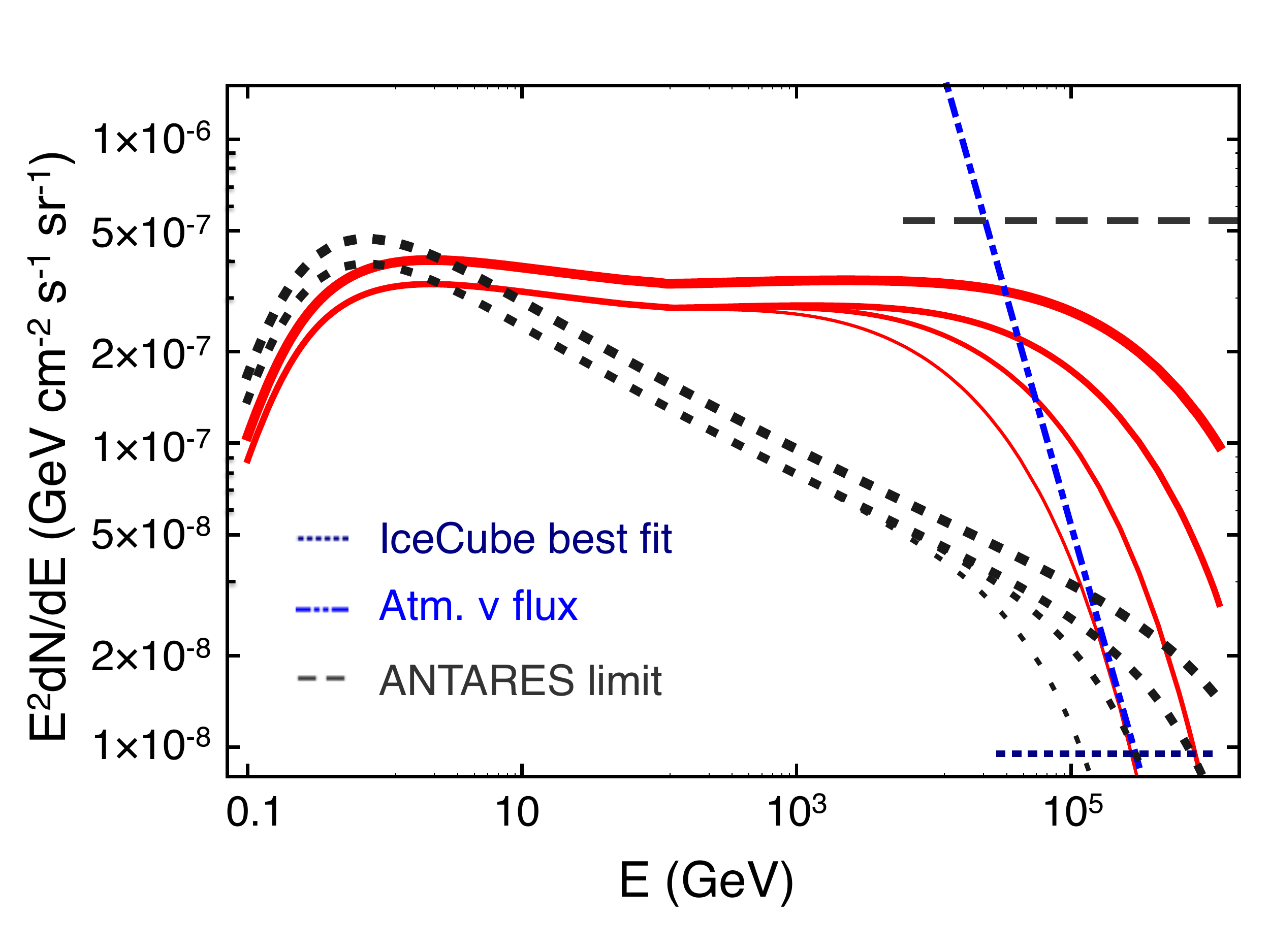}
\caption{The pre-oscillated $\numu$ and $\barnumu$ flux from the FB, as a function of the energy, normalized to the gamma ray flux, for spectral indices $k=2.1$ (solid, red) and $k=2.3$ (dotted, black). For each index, the curves from thick to thin correspond to $E_0=30,10,3,1$ PeV. The dash-dot curve presents the atmospheric \n\ flux \cite{Honda:2006qj} averaged over $25^\circ$-$95^\circ$ zenith angle. The dotted line is the ANTARES upper limit \cite{Adrian-Martinez:2013xda}. The dashed line is the estimated \ic\ best-fit diffuse flux expected for the FB region \cite{Aartsen:2013jdh}. This figure has been taken from Ref.~\cite{Lunardini:2013gva}}
 \label{fluxfig}
\end{figure}

We adopt the FB neutrino fluxes from Ref.~\cite{Lunardini:2011br}, to calculate the expected event rate at \ic\ due to FB . These neutrino fluxes are produced from the interactions of cosmic ray protons from supernova remnants with the ambient gas. The proton spectrum has the form of $dN/dE\propto E^{-k} \exp(-E/E_0)$, where $E_0$ is the cutoff energy, motivated by the maximum energy that accelerated protons may reach, varying from 1 - 100 PeV \cite{Ptuskin:2010zn}. By fitting the gamma ray data of the FB, the spectral index $k=2.1$ is taken as our default model; meanwhile, due to the limited number of gamma ray data points, a steeper spectrum with $k=2.3$ is also compatible with observation, as seen in Fig.~\ref{fluxfig}. These two sets of fluxes depending on $E_0$ differ significantly above $\sim 200$ GeV. The flux with cutoff energy of $E_0=30~PeV$ is the most optimistic one, which is approximately $20\%$ higher than the normalization of the whole flux, allowed by the uncertainty in the gamma ray data. The \ic\ best-fit astrophysical flux of $E^2dN/dE=0.95 \times 10^{-8}$ GeV/cm$^2$/s/sr is shown as well \cite{Aartsen:2014gkd}, see Fig.~\ref{fluxfig}. 

The neutrino fluxes initially produced at the FB of all flavors are in a composition of $\numu$: $\nue$ : $\nutau$ $= 2 : 1 : 0$. After oscillations, the flavor ratios are close to $\nue$ : $\numu$ : $\nutau$=$ 1 : 1 : 1$, with deviations up to $\sim 30\%$ at $E \sim 1$ PeV.

The atmospheric muons and neutrinos are the two main backgrounds at \ic\ , with an expectation of 8.4 and 6.6 events respectively for a 988-day running time \cite{Aartsen:2014gkd}. The atmospheric muon level is depending on the detector veto, more details in \cite{Aartsen:2013jdh}. For the atmospheric neutrinos, we adopt the $\nu_{\mu}$ flux model from Ref. \cite{Honda:2006qj}, which is a good fit of \ic's atmospheric data \cite{Abbasi:2010ie}, extrapolated at high energies. In our calculation, we consider the flux to be symmetric in $\cos \theta_z$ \cite{Athar:2012it}, take the $\numu/\nue$ ratio of 14 \cite{Sinegovskaya:2013wgm}, and neglect flavor oscillation due to the short propagation distance \cite{Gaisser:1997eu}, as shown in Fig.~\ref{fluxfig}. 

\section{Event Rate}

\begin{figure}[htbp]
\centering
\includegraphics[width=0.4\textwidth]{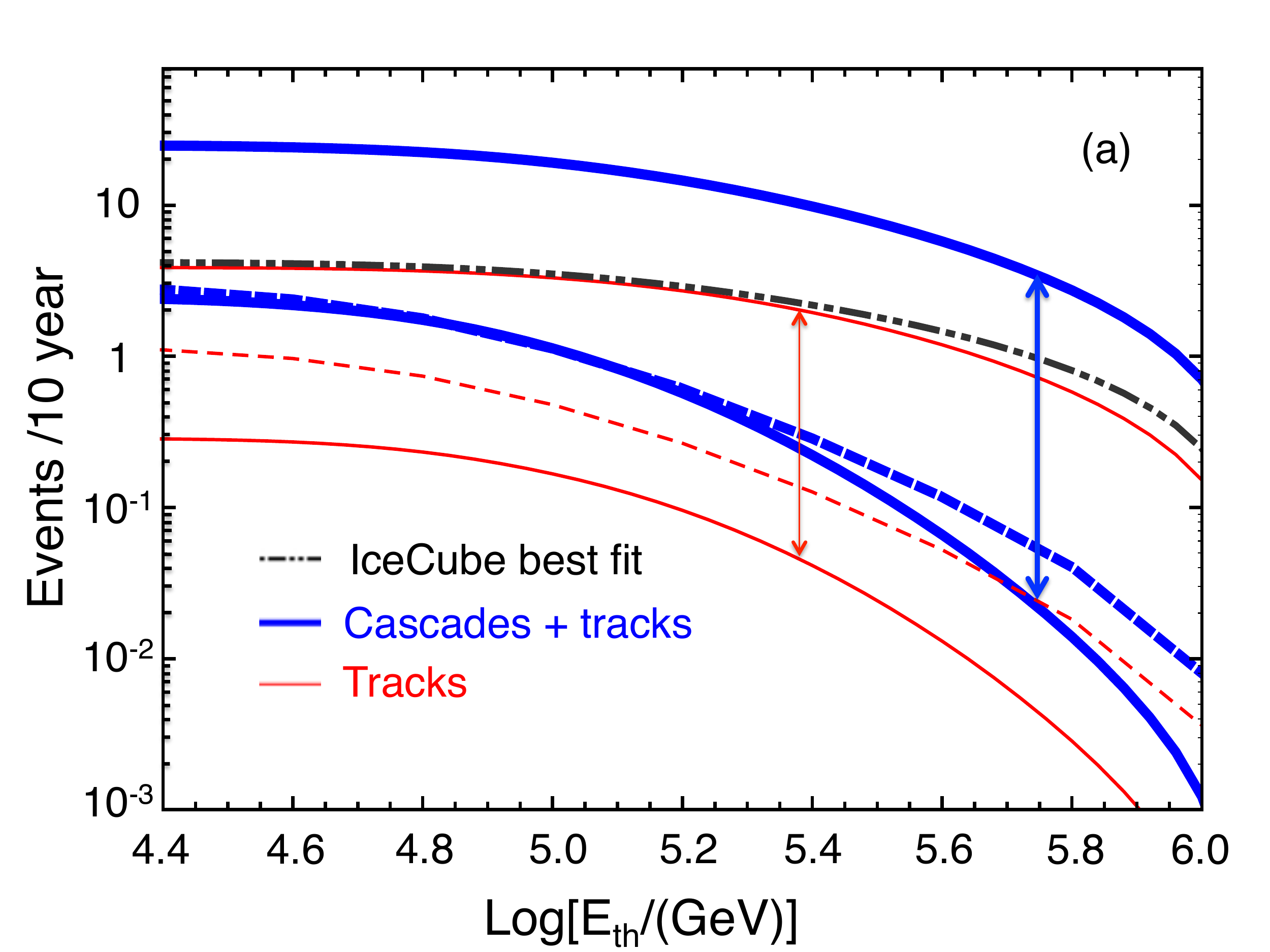}
\includegraphics[width=0.4\textwidth]{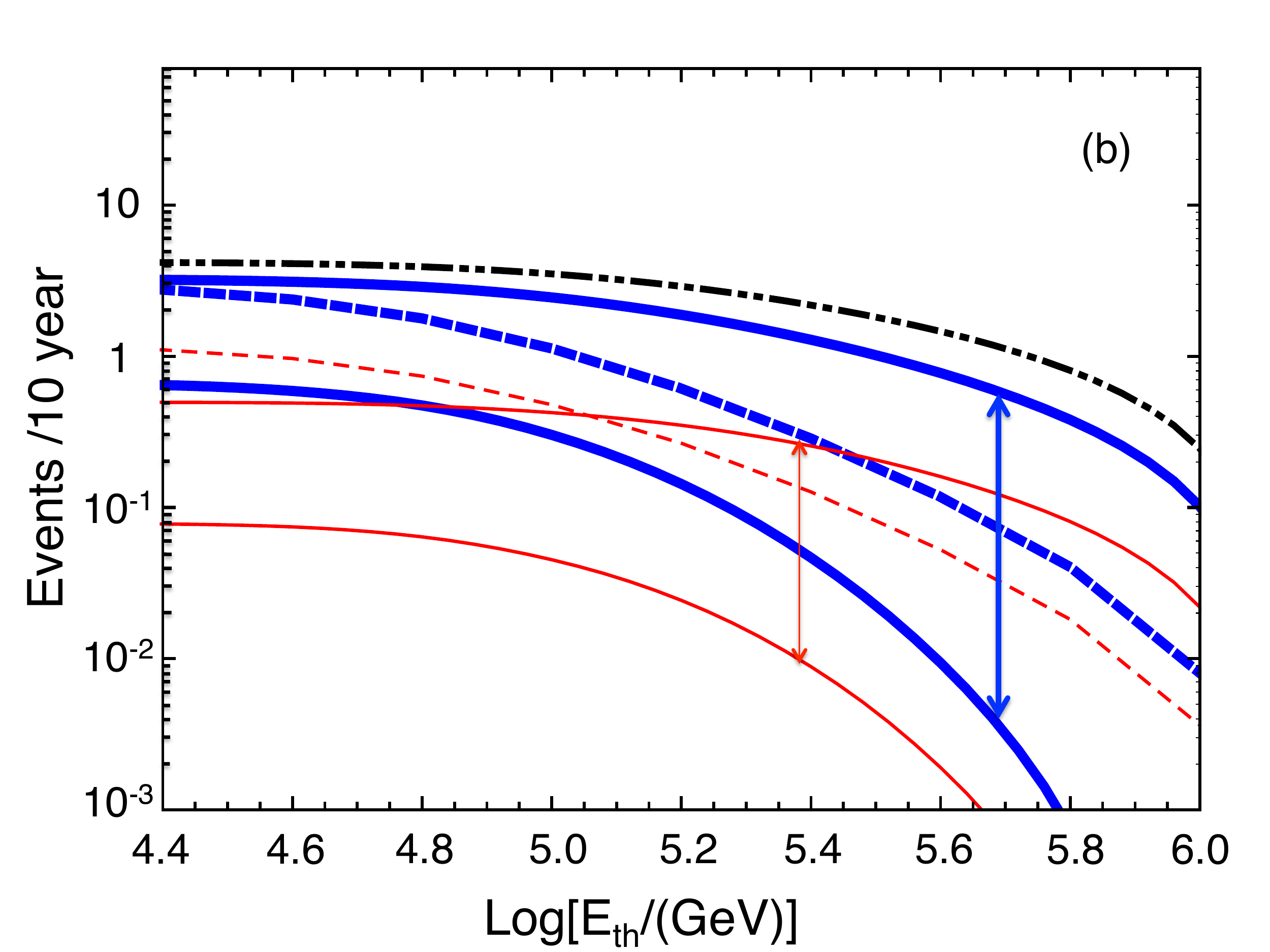}
 \caption{Expected number of events at IceCube for 10 years, as a function of energy threshold $E_{th}$, for the spectral indices of $k=2.1$ (a) and $k=2.3$ (b), from Ref.~\cite{Lunardini:2013gva}. The total FB (atmospheric neutrino) signals of showerlike and tracklike events are presented as thick solid (dashed) curves. The tracklike FB (atmospheric neutrinos) signals are thin solid (dashed) curves. The arrows indicate the variation of the primary spectrum cutoff in the interval $E_0 = 1 - 30$ PeV. The estimated total showerlike and tracklike events from \ic\ best-fit diffuse flux.}
\label{ratefig}
\end{figure}

The expected number of atmospheric neutrino backgrounds and signals for $k=2.1, 2.3$ in 10 years, above the energy threshold $E_{th}$ is estimated, as shown in Fig.~\ref{ratefig}. Taking into account of the angular resolution $1^\circ$ and $10^\circ$ for tracks and showers respectively at \ic\ \cite{Aartsen:2013jdh}, we calculate the number of atmospheric shower events over a larger solid angle than that of FB (0.808 sr). To do this, we encase each bubble in a rectangle in the $\theta$ and $\phi$ coordinates (see Fig.~1), and enlarge the area by $15^\circ$ on each side. On the other hand, we take FB solid angle for track events, due to the angular resolution less than $1^\circ$. As seen in Fig.~4, the atmospheric shower (thick, dashed) and track (thin, dashed) events are comparable. This can be explained that the predominant atmospheric $\numu$ flux compensates for the smaller effective area for tracks. For a harder spectrum ($k=2.1$), the signal is above the background with cutoff energy $E_0 \gta  10~{\rm PeV}$. With $E_0 = 30~{\rm PeV}$, the signal is approximately 2 orders of magnitude higher. For $E_{th}=40 ~{\rm TeV}$, we find 23 signals and 3 background events, indicating a $\sim 4~\sigma$ excess due to the FB. It would be quite promising to discover the FB with the use of detailed statistical analyses of the spatial correlation. For $E_0 \sim 1~{\rm PeV}$, the signal is even lower than the background for some thresholds, and with lower event rate as well. For a steeper spectrum ($k=2.3$), even with $E_0 = 30~{\rm PeV}$, the signal is still not significant, comparable to the background.

\begin{figure}[htbp]
\centering
\includegraphics[width=0.4\textwidth]{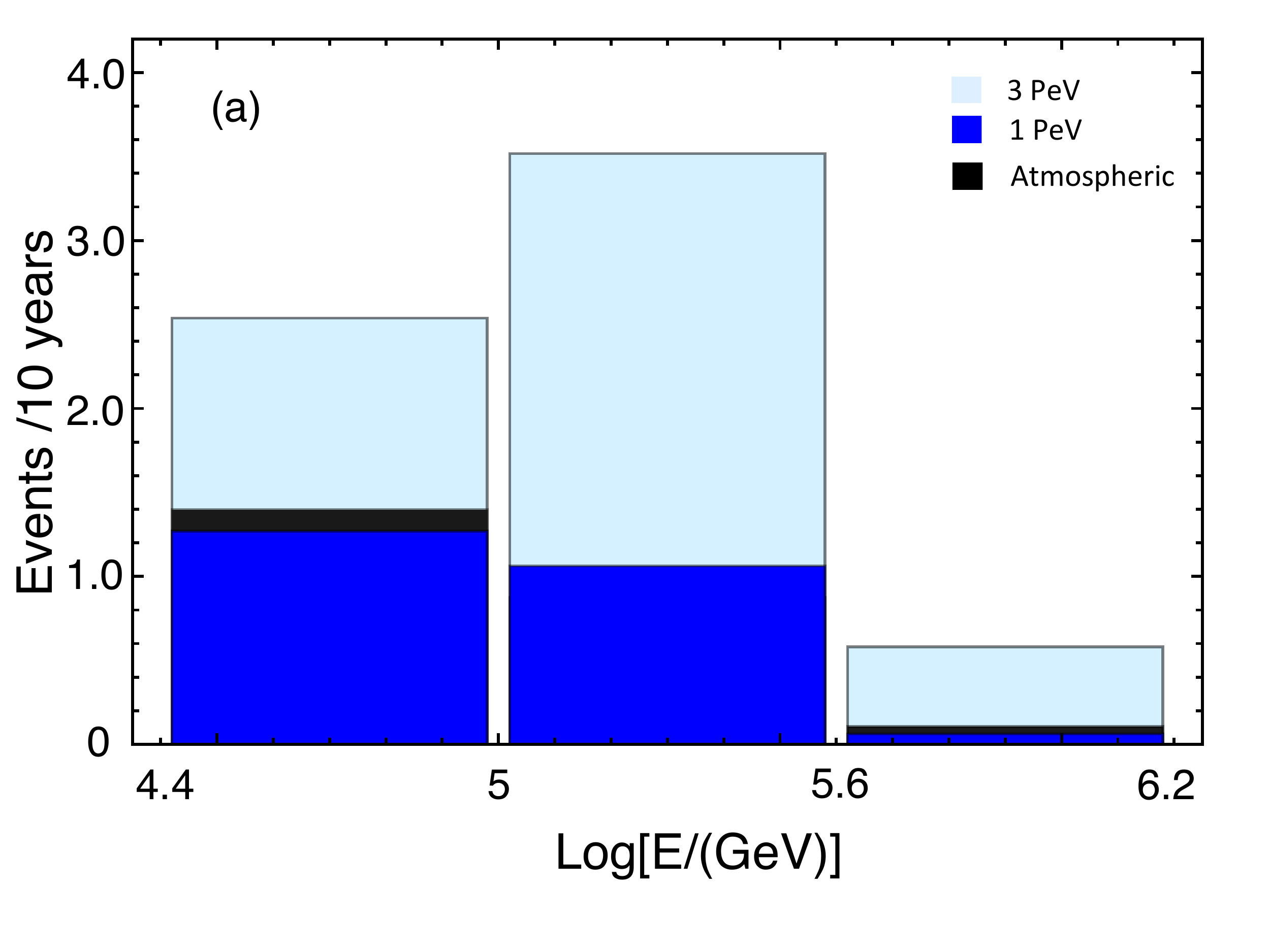}
\includegraphics[width=0.4\textwidth]{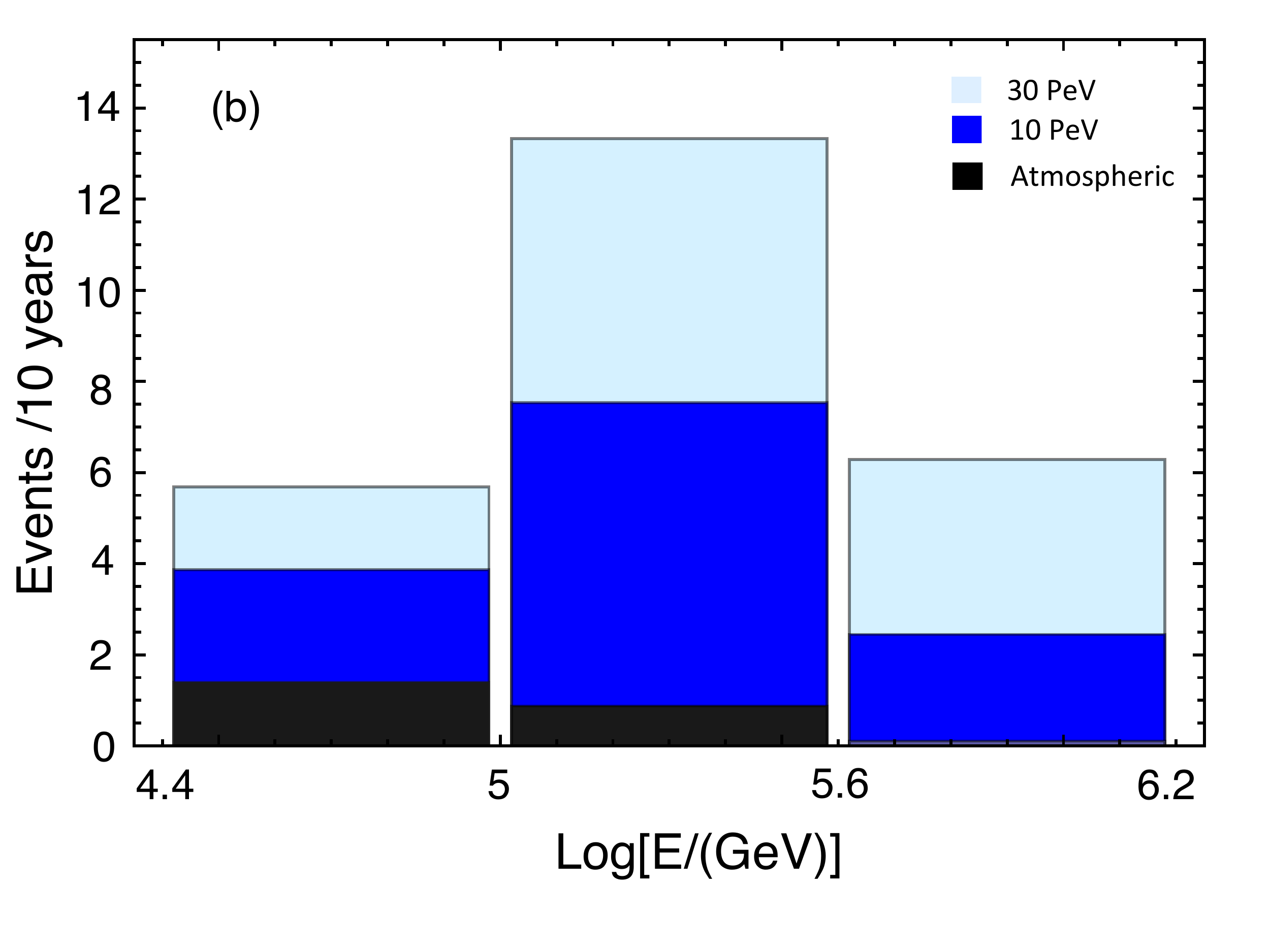}
\caption{The expected number distribution of shower and track events per decade in each energy bin for atmospheric background and signal with $k=2.1$ and cutoff energy $E_{0}=1, 3$ PeV (a) and $E_{0}=10, 30$ PeV (b), from Ref.~\cite{Lunardini:2013gva}.}
\label{histofig}
\end{figure}

The distribution of number of signal and background events per decade in neutrino energy bins is given in Fig.~\ref{histofig}. As confirmed from Fig.~4(a), for $E_{0}=1$ PeV, the signal is comparable to the background, and the most events are distributed in the energy of $10^{5} - 10^{5.6}$ GeV due to the rise in the effective area below $\sim 1~{\rm PeV}$ \cite{Aartsen:2013jdh}. 

\begin{table}
\caption{The expected number of shower and track events from the atmospheric background and from the \fb\ in three neutrino energy bins, with different cutoff energy $E_0$, for a 988-day \ic\ livetime. The number of tracklike events are in brackets. }
\begin{center}
\begin{tabular}{ccccc}
\br
 $E~ ({\rm GeV})$ & $10^{4.4} -10^{5} $ & $10^5 - 10^{5.6} $  & $10^{5.6} - 10^{6.2} $ & Total \\
\mr
$E_0=1 {\rm PeV}$ &  0.34 & 0.29 & 0.02 & 0.65 \\
                         &   [0.03] & [0.04] & [0] & [0.08] \\
$E_0=3 {\rm PeV}$  &  0.69 & 0.95 & 0.16 & 1.8 \\
                         &   [0.07] & [0.14] & [0.03] & [0.24]\\
$E_0=10 {\rm PeV}$  & 1.05 & 2.04 & 0.66 & 3.75 \\
                          &  [0.1] & [0.32] & [0.14] & [0.56] \\
                          
$E_0=30 {\rm PeV}$  &   1.54 & 3.61 & 1.7 & 6.85 \\
                            & [0.15] & [0.57] & [0.35] & [1.08] \\
                            
 Background & 0.38 & 0.24 & 0.03 & 0.64 \\
                            &   [0.11] & [0.09] & [0.01] & [0.21] \\
\br
\end{tabular}
\end{center}
\label{tabrate}
\end{table}%

\begin{figure}[htbp]
\centering
\includegraphics[width=0.5\textwidth]{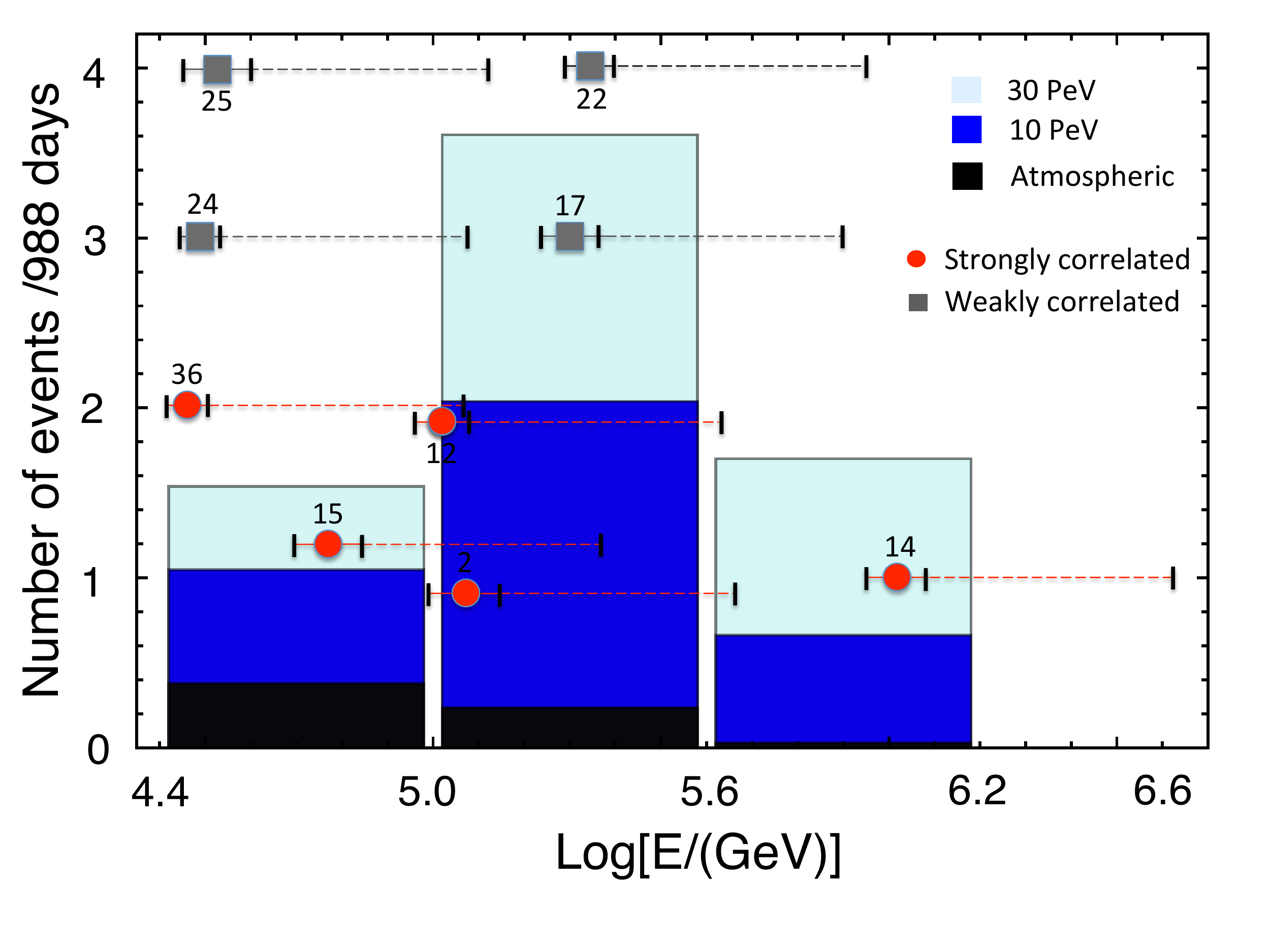}
\caption{The same as Fig.~\ref{histofig}(b) for the \ic\ 988-day running time. Four strongly (red dot) and four weakly (grey square) correlated with the \fb\ events at \ic\ are shown. The solid and dashed error bars present the errors on the observed energy and a factor of $\sim 3-4$ difference between neutrino energy and deposited energy for neutral current events \cite{Aartsen:2013vja}, respectively. Their coordinates on the vertical axis present the number of events for which the central value of the observed energy falls in the corresponding bins. }
\label{ice3rate}
\end{figure}

For comparison with \ic\ data, we rescale the event rate with a 988-day running time \cite{Aartsen:2014gkd}, as seen in Table \ref{tabrate} and Fig. \ref{ice3rate}. The total shower and track background events is less than one in the energy of $10^{4.4} - 10^{6.2}$ GeV. For $E_0>3$ PeV, it is expected more than 2 signals from FB with a livetime of 988 days. Especially, for $E_0=30$ PeV, $N\sim 5$ and $N\sim 2$ events are estimated below and above $E=10^{5.6}~{\rm GeV} \simeq 400~{\rm TeV}$ of neutrino energy respectively. Which is intriguingly close to the number of 5 events that has strongly spatial correlation to the FB.

\section{Discussion}

In this work, we present the possibility that \ic\ may have detected the FB. With a hard spectrum and the most optimistic neutrino flux model $E_0=30$ PeV, we predict that up to 5 of \ic\ observed events may due to the FB. Especially, 9 \ic\ events are spatially correlated with the FB (Fig. \ref{icdata}), which might be the conservative upper limit for the number of events. To have a significant detection of the FB, the statistics not only depends on the backgrounds, but also on the level of the other neutrinos sources. However, as seen in Fig. \ref{ratefig}, if with the most promising flux model, in seven to ten running years, the signals will be well identified. 

The observation of a neutrino flux from the FB will provide clues to the mechanism of the FB, to the maximum limit of particle acceleration in supernova remnants, and to the time scale of the activity of the Galactic center as well.

\ack
We thank Albrecth Karle, Mariola Lesiak-Bzdak, Jakob van Santen and
Nathan Whitehorn.  C.L. and L.Y. acknowledge the
National Science Foundation grant number PHY-1205745.
K.T. acknowledges the ASU/NASA Space Grant  2013 for partial support. S.R. acknowledges support from the National Research Foundation (South Africa) grants CPRR 2014 number 87823 and BS 2014 number 91802.

\end{document}